\documentclass[twocolumn,10pt,aps,superscriptaddress,pra]{revtex4-1}
\usepackage{epsfig}
\usepackage{amsmath}%
\usepackage{amsfonts}%
\usepackage{amssymb}%
\usepackage{graphicx}
\begin{document}
\bibstyle{plain}

\title{Decoherence through Spin Chains: Toy Model}
\author{Marcin Wie\'sniak}\affiliation{Institute of Theoretical Physics and Astrophysics, University of Gda\'nsk, PL-80-952 Gda\'nsk, Poland}\affiliation{Institute for Experimental Physics, University of Vienna, A-1090 Vienna, Austria}
\begin{abstract}
The description of the dynamics of closed quantum systems, governed by the Schr\"odinger equation at first sight seems incompatible with the Lindblad equation describing open ones. By analyzing closed dynamics of a spin-$\frac{1}{2}$ chain we reconstruct exponential decays characteristic for the latter model. We identify all necessary ingredients to efficiently model this behavior, such as an infinitely large environment and the coupling to the system weak in comparison to the internal couplings in the bath.
\end{abstract}
\maketitle
Quantum mechanics is distinguished from other probabilistic theories by the description of the relations between various observables. These relations are described by coherences, off-diagonal entries of a matrix representing a state. This fact has important consequences for dynamics of quantum systems.

Understanding this dynamics has given rise to many important applications, such as parametric down-conversion \cite{spdc}, perfect state transfer \cite{pst}, or sudden death \cite{suddendeath} (rebirth \cite{rebirth}, etc.) of entanglement. However this evolution might be seen at many different levels. The first is statical. One just applies a completely positive map to an initial state to get the final one \cite{Kraus1,Kraus2,Choi}. This is a black box approach, where we are not interested in underlying physical phenomena. The second level is to solve the Schr\"odinger equation \cite{schroed} to be able to trace the evolution of the system at any given instance of time. We assume here, however, that a quantum system is closed, which is often clearly false in light of experimental data. We can, of course, incorporate the environment, the ``invisible'' part of the system into the equation, which most often greatly complicates its solutions. 

The third description of dynamics is based on the master equation \cite{lindblad}. It has terms describing the free evolution governed by the Schr\"odinger equation and some phenomenological terms describing the interaction with the environment.   

While the completely positive map approach is general, the Schr\"odinger equation and the master equation seem incompatible. The former generates a unitary evolution group. This causes the probability of a survival of the initial state to diminish quadratically with time (for short times). Consequently, we observe the so-called quantum Zeno effect \cite{Zeno1,Zeno2}. If we frequently measure in the basis containing the initial state, the free dynamics of the system freezes -  the survival probability of the initial state tends to 1 in the limit of continuously repeated measurements. Moreover, while the free evolution is reversible and even in time, the master equation describes exponential decays. The reverse evolution of an open system might make the state unphysical. How can these descriptions be agreed with each other?

Here we want to show, how exponential decays of coherence might follow from the Schr\"odinger equation. The evidence will be provided by a simple system, the semi-infinite chain of qubits coupled by the $xx$ interaction. We will utilize the Heisenberg picture formalism \cite{Heis1,Heis2,Heis3}, which turned out to be successful in analyzing the state initiation-free perfect state transfer \cite{Heis4,Heis5} and Hamiltonian tomography with limited access \cite{Heis6,Heis7}. 

There have been already attempts to simulate decoherence with a collection of spins, for example \cite{Wang}. It should be pointed out, however, that Wang, Wang, and Su, as well as many other Authors \cite{decoh1,decoh2,decoh3,decoh4,decoh5,decoh6,decoh7,decoh8,decoh9,decoh10,decoh11}, simulate the environment with a finite number of spins. In systems of finite size there is no possibility to have no remains of the initial state after infinitely long times. One may even expect a revival of the initial state with an arbitrary fidelity due to Poincar\'e recurrences \cite{pastawski1,pastawski2}, however rare (see Figure 1). The deficits are removed in our approach.

A detailed analysis of infinite spin chain baths, though in a slightly different context, in Refs. \cite{pastawski3,pastawski4}. Ref. \cite{pastawski3} reconstructs the exponential decay in the approximate calculations, while Ref. \cite{pastawski4} studies the behavior of the magnetization of one of the sites. These results are complementary to ours, which obtained within a different formalism.
\begin{figure}
\centering
\includegraphics[width=6cm]{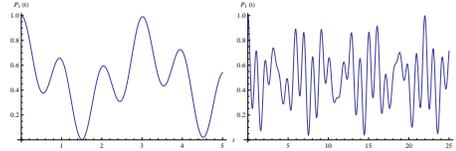}
\caption{(Color Online) In an imaginary system, whose evolution is governed by only two mutually irrational frequencies, $1,\pi$, the survival probability $P_1=\frac{1}{4}(2+\cos 2t+\cos 2\pi t)$ (left panel) never reaches 1 for $t>0$ and has a peak above 0.9 much earlier than $P_2=\frac{1}{6}(3+\cos 2t+\cos 2\pi t+\cos 2 e t)$ corresponding to a system with three mutually irrational frequencies,$1,\pi$, and $e$ (right panel). Only a system with infinitely many such frequencies allow (but does not guaranty) that such peaks never occur.}
\end{figure}

The system of our interest will be just a single qubit, labeled with 0. The role of the environment will be played by a wire of spins-$\frac{1}{2}$, labeled 1 to $\infty$. Denote Pauli matrices acting on $n$th qubit by $X_n, Y_n,$ and $Z_n$.

The overall Hamiltonian is
\begin{eqnarray}
H&=&H^{free}+H^{plug}+H^{wire}\nonumber\\
&=&0+\frac{K_0}{2}(X_0X_1+Y_0Y_1)+\frac{K}{2}\sum_{i=1}^\infty(X_{i}X_{i+1}+Y_{i}Y_{i+1}),\nonumber\\
\end{eqnarray}
where we have taken the system of interest to be free, and given name ``plug'' to the interaction between the system and the environment (in analogy to ``wire'', as the chain is called). We can now take the initial state of the 0th spin to be $\rho=\frac{1}2(1+\vec{v}\cdot(X_0,Y_0,Z_0))$, and the maximally mixed state as the state of the wire. In such a case, the evolution of $A=X_0,Y_0,Z_0$ and is given by
\begin{eqnarray}
A(t)&=&\sum_{j=0}^\infty\frac{(it)^j}{j!}L^j(A)
\end{eqnarray}
with $L(\cdot)=[H,\cdot]$. In particular, we have
\begin{equation}
X_{0}(t)+\alpha_0(t)X_{0}+\alpha_1(t)Z_{0}Y_{0}+\alpha_2(t)Z_0Z_1X_2+...\!
\end{equation}
(confirm  Ref. \cite{paternostro}). Alternatively, we can write the evolution in the matrix form:
\begin{eqnarray}
\left(\begin{array}{c}X_0\\Z_0Y_1\\Z_0Z_1X_1\\...\end{array}\right)(t)&=&\exp\left[ t\left(\begin{array}{cccc}
0&K_0&0&...\\
-K_0&0&-K&...\\
0&K&0&...\\
...&...&...&...
\end{array}\right)\right]\nonumber\\
&\times&\left(\begin{array}{c}X_0\\Z_0Y_1\\Z_0Z_1X_1\\...\end{array}\right).
\end{eqnarray}
The problem can be hence mapped onto a classical random walk on the set of non-negative numbers. In our problem we observe only the 0th spin, hence we are interested in the number of $n$-step ($n$ being even) walks which start and end at 0. Since in general $J_0\neq J$, we need to categorize these walks in terms of how many times they pass the initial position (not including the start and the end, we call this number $k$). The numbers of such walks $l(n,k)$ are given in Table 1.
\begin{table}
\begin{tabular}{|c|c|c|c|c|c|c|}
\hline
&\multicolumn{6}{|c|}{k}\\
\hline
n&0&1&2&3&4&5\\
\hline
2&1&0&0&0&0&0\\
\hline
4&1&1&0&0&0&0\\
\hline
6&2&2&1&0&0&0\\
\hline
8&5&5&3&1&0&0\\
\hline
10&14&14&9&4&1&0\\
\hline
12&42&42&28&14&5&1\\
\hline
\end{tabular}
\caption{Number of $n$-step walks on non-negative numbers starting and ending at 0, and going through the 0th position $k$ times.}
\end{table}
Collectively, these quantities are described by
\begin{equation}
l(n,k)=\frac{(k+1)(n-k-2)!}{\left(\frac{n}{2}-k-1\right)!\frac{n}{2}!}
\end{equation}
for $2k\leq n$. Thus the original operators in the Heisenberg picture satisfy
\begin{eqnarray}
\label{evolution}
\alpha_0(t)&=&tr(X_0(t)X_0(0))=tr(Y_0(t)Y_0(0))\nonumber\\
&=&\sum_{j=0}^\infty(-1)^j\frac{t^{2j}}{(2j)!}\sum_{k=0}^{2j}K_0^{2k+2}K^{2(j-k-1)}l(2j,2k)\nonumber\\
&=&1+\sum_{j=1}^\infty\left((-1)^j\frac{(Kt)^{2j}}{(2j)!}\frac{K_0}{K}(2j-2)!\right.\nonumber\\
&\times&\left.\frac{\,_2F_1\left(1-j,2,2-2j,\frac{K_0}{K}\right)}{(j-1)!j!}\right),\nonumber\\
&&tr(Z_0(t)Z_0(0))=tr(X_0(t)X_0(0))^2.
\end{eqnarray}
Notice that the state of infinitively many qubits cannot be formally written as it faces the normalization problem. We have conveniently avoided this difficulty by studying the auto-fidelity of operators, $-1\leq tr(A(t)A(0))\leq 1$.

The remaining part of the paper will be devoted to analyzing how a general state of a qubit, $\rho=\frac{1}{2}(1+v_x X_0+v_y Y_0+v_z Z_0)$ behaves under the evolution of the whole chain.
 
The first, trivial, case is of $K=0$. Then the ``environment'' contains only one qubit. We have $\alpha_0(t)=\cos K_0t$. As expected, information contained in the system revives periodically, since it has nowhere to go.

For $K\neq 0$, there are very few analytical solutions. One case is $K=\frac{K_0}{\sqrt{2}}$, where we have
\begin{equation}
\alpha_0(t)=1-\frac{(Kt)^2}{2}+\frac{(Kt)^4}{4}-\frac{(Kt)^6}{36}+...=J_0(K t).
\end{equation}

Note hat the limit value of the Bessel function for its argument going to $\infty$ is $0$. Hence, in the limit of long time the is no information at the 0th site about its initial state. It can be even argued 	that this information can be found nowhere specifically in the chain -- it is completely delocalized. We already see how we benefit from considering an infinitively high-dimensional environment.

It is known from the theory or spin chains that the $K_0=\sqrt{2}K$ situation corresponds to an infinite spin chain, with the index ranging from $-\infty$ to $\infty$. In such a case it is possible to analyze other phenomena. For example, Figure 2 presents the length of the Bloch vector of the zeroth qubit, when initially it was in state $|+\rangle\langle +|=\frac{1}{2}(1+X_0)$, wheres the environment was fully magnetized (each spin of it in state $|0\rangle\langle 0|=\frac{1}{2}(1+Z_i)$). Notice that the system periodically becomes completely polarized.

\begin{figure}
\centering
\includegraphics[width=6cm]{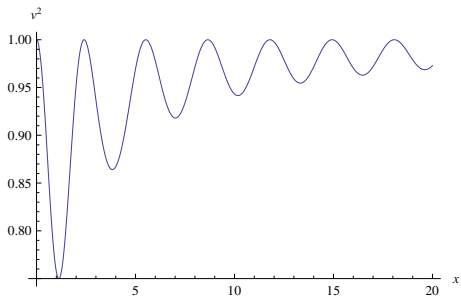}
\caption{(Color Online) Evolution of the length of the Bloch vector $v^2=v_x^2+v_y^2+v_z^2$ of the 0th spin with $K_0=\sqrt{2}K$ the initial state of the chain being $|+000...\rangle$. $v^2=1$ corresponds to a completely magnetized state.}
\end{figure}
Note that the relevant experimental results presented in \cite{science} are in high agreement with our analysis, even though the number of parallel waveguides was only 21 in the experiment.

The other analytical solution is for $K=K_0$, for which we have
\begin{equation}
\alpha_0(t)=1-\frac{(Kt)^2}{2}+\frac{(Kt)^4}{12}-\frac{(Kt)^6}{144}+...=\frac{J_1(2Kt)}{Kt}.	
\end{equation}
Notice that this function also vanishes for $t\rightarrow\infty$, but its envelope is tighter than in the previous case. Instead of $\frac{1}{\sqrt{t}}$, it now behaves like $t^{-\frac{3}{2}}$ for long times. It is a general trend: as we increase ratio $\frac{K}{K_0}$, the decay of $\alpha_0$ will be faster. The solution for $K=K_0$ will be referred to in the later considerations.

Further, not much can be said analytically. One thing we can show is that the first inflection point as we fix the characteristic time of the decay, $\frac{K}{K_0^2}$, but go with $\frac{K}{K_0}$, which characterizes the decay of information contained in the $0$th spin, to infinity, we can show that the first inflection point tends of $\alpha_0(t)$ to $0$. Namely, if one truncates $\partial_x^2\alpha_0(\frac{K}{K_0^2}x)$ to the second order
\begin{equation}
-\frac{K^2}{K^2_0}+\frac{K^4(K^2+K_0^2)}{2K_0^6}x^2=0,
\end{equation}
which has solutions at $x_0=\pm\sqrt{2}\left(\frac{K^2}{K^2_0}+\frac{K^4}{K^4_0}\right)^{-\frac{1}{2}}$, clearly tending to 0 in the limit. This suggests that the quadratic approximation is relevant for shorter and shorter times.

Figure 3 shows the following quantity:
\begin{equation}
\label{measure}
\chi=\int_0^1\left(\alpha'_0(x)-e^{-x}\right)^2,
\end{equation}
in function of $\frac{K}{K_0}$. $\alpha'_0(x)$ is $\alpha_0(t)$ truncated up to 20th power of $x$. From this plot and Figure 4, which shows $\alpha_0(t)$ for various ratios $\frac{K}{K_0}$, we conclude that in the limit of very high $K$ the decay of information in the 0th system becomes exponential. We must fail, however, to demonstrate it analytically. Since the Schr\"odinger equation gives only evolution even in time, the limit is $\exp\left(-\frac{|t|J^2}{J_0}\right)$, and the function has a point of non-smoothness at $t=0$. We have $\alpha_0(t)$ only as a Maclaurin series, and hence we necessarily end up with a non-analytical function.

\begin{figure}
\includegraphics[width=6cm]{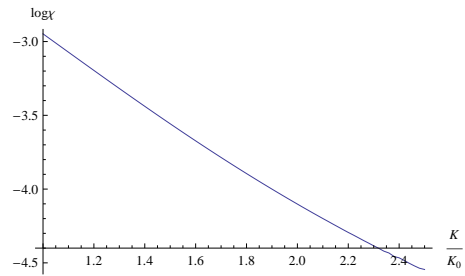}
\caption{(Color Online) Logarithm of $\chi$ given by Eq. (\ref{measure}) in function of $\frac{K}{K_0}$.}
\end{figure}

\begin{figure}
\includegraphics[width=6cm]{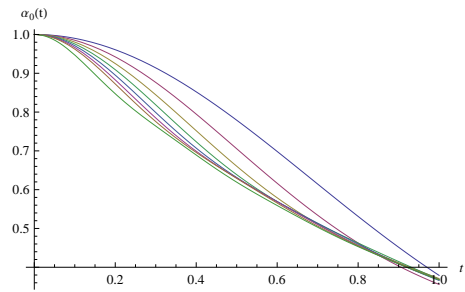}
\caption{(Color Online) $\alpha_0(t)$ in function of time. The curves from the highest to the lowest at $t=0.2$ are plotted for $\frac{K}{K_0}=\sqrt{2},\sqrt{3},2,\sqrt{5},\sqrt{6},\sqrt{7},2\sqrt{2}$ and $2\sqrt{3}$. $\frac{K}{K^2_0}$ has been fixed to 1.}
\end{figure}
We can only describe the decay in a phenomenological way. Let us use the Trotter decomposition to the evolution,
\begin{eqnarray}
\exp(iHt)&=&\lim_{n\rightarrow\infty}\left(\exp\left(\frac{itH_{free}}{n}\right)\exp\left(\frac{itH_{plug}}{n}\right)\right.\nonumber\\
&\times&\left.\exp\left(\frac{itH_{wire}}{n}\right)\right)^n.
\end{eqnarray}  
We start with the state of the 0th qubit $\rho=\frac{1}{2}(1+X_0)$. The plug Hamiltonian changes operator $X_0$ into $Z_0Y_1$ in some part. As we switch off the plug interaction and switch on the wire interaction, it instantly delocalizes $Z_0Y_1$, practically mapping it to 0. In this approach we are unable to derive the lifetime of information, and we again face the problem of the Zeno effect. However, the obtained decays are clearly exponential. This allows us to speculate about more general models of decoherence.

The specific Hamiltonian of the ``wire'' is in fact irrelevant. The only thing that matters is that it maps any localized operator to an unlocalized one in an infinitely short time. Hence, two features of the environment are relevant. First that they are infinitely dimensional, and that the coupling constants between its constituents are scale properly with respect to the ``plug'' coupling constants.

The choice of a specific plug determines the kind of decoherence. The main question is whether there exists an orbit of operators acting on the system, which commutes with the plug. For example, we could have taken the Ising interaction, $H_{plug}=Z_0Z_1$, rather than the {\em xx} interaction, we see that the resulting map preserves the $z$ component of the Bloch vector for the 0th spin (unless there is a transverse magnetic field acting on the system). In this way we easily see that the ``pancake map'', which maps a Bloch vector $(v_x,v_y,v_z)$ to $(v_x,v_y,0)$, is unphysical. For this observation, we do not need to refer to the concept of the complete positivity, but only to the non-existence of a plug, which commutes with $X_0$ and $Y_0$, but not with $Z_0$. If from operators commuting with the plug we can build a Lie algebra, there exists a decoherence-free subspace.

Of course, the toy model presented here allows to investigate more phenomena related to decoherence. If our system consists of two qubits, each $xx$-coupled to a separate wire, in the limit of $\frac{K}{K_0}\rightarrow\infty$, and we initialize the system in the singlet state, we observe that entanglement dies out after a finite time \cite{suddendeath}. If, however, we choose $K_0$ sufficiently high in relation to $K$, $\alpha_0(t)$ becomes oscillating. We still may have the sum of squares of correlations between the two qubits, $\sum_{i,j=1}^3T_{ij}^2$ greater than $1$, indicating entanglement \cite{expfrie}. We hence observe the so-called rebirth of entanglement \cite{rebirth}, occurring in non-Markovian environments, such as high-Q optical cavities.

In conclusion, we have argued that a semi-infinite chain of spins-$\frac{1}{2}$ can serve as a proper toy-model for decoherence. We have given the numerical evidence that in the properly taken limit of the weak system-environment coupling, exponential decays of information are observed. Moreover, due to a well established formalism of dynamics of spin-$\frac{1}{2}$ chains, we have been also able to study some non-Markovian cases. We have captured all the necessary ingredients for decoherence (the environment of an infinite dimension and of an infinitely fast internal dynamics) and discussed phenomena, such as the sudden death and rebirth of entanglement, or decoherence-free subspace.

An important issue is whether the state of of the ``wire'' changes during the evolution. Of course, the unitary evolution preserves the amount of information contained in the whole system. As it leaks out from the 0th qubit, it appears in the ``wire''. However, it would take a simultaneous observation of infinitively many systems to recover this information, hence we can take the state of the wire effectively as maximally mixed.    

Interestingly, in the weak coupling limit we observe finite decay times for infinite values of $K_0$ and $K$.

Our result provokes some open questions. For example, we have conveniently assumed that spins in the environment are already in a maximally mixed states. It shall not be difficult to perform a similar analysis for a completely magnetized environment, but can one efficiently model finite temperatures by biased random orienting spins? This would allow to 	generate mixed states of the system and the environment, while keeping the ``universe'' in a pure state. Then again, there would be a need to explain mechanisms behind this random orientation. Another question is whether there are any differences when a Heisenberg chain of spins-1, not an $xx$ chain of spins-$\frac{1}{2}$, is used. The former kind has a gap between the ground state and the first excited state, even in the thermodynamical limit.

{\em Acknowledgements:} The research was supported by the ERC grant (QIT4QAD) and the program ``Optimization of Quantum Resources'' of the Ministry of Science and Higher Education of Poland. The Author greatfully acknowledges Mio Murao for comments on the project and Horatio Pastawski for pointing out relevant references.

\end{document}